%% file: No_fly_zone_V1release.tex
\documentclass[letterpaper,journal]{IEEEtran}
\usepackage{amsthm, amssymb, graphicx, multirow,amsmath, color, amsfonts}
\usepackage{optidef}
\usepackage[update,prepend]{epstopdf}
\usepackage[noadjust]{cite}
\usepackage[latin1]{inputenc}
\usepackage{tikz}
\usetikzlibrary{arrows,calc}		
\usepackage{bbm}
\usepackage{bm}
\usepackage{pdfpages}
\usepackage{amsmath}
\usepackage{tabularx}
\usepackage{multirow}
\usepackage{subfigure}
\usepackage{comment}
\usepackage{algorithm}
\usepackage{algpseudocode}

\algnewcommand\algorithmicinput{\textbf{INPUT: }}
\algnewcommand\Input{\item[\algorithmicinput]}
\algnewcommand\algorithmicoutput{\textbf{OUTPUT: }}
\algnewcommand\Output{\item[\algorithmicoutput]}
\input{notation}

\allowdisplaybreaks 

\begin{document}
\graphicspath{{./Figures/}}
\title{ Optimal No-Fly Zone Design for the Coexistence of Drone and Satellite Networks
}
\author{ Xiangliu Tu, Chiranjib Saha,~\IEEEmembership{Member,~IEEE},  Harpreet S. Dhillon,~\IEEEmembership{Fellow,~IEEE} 

\thanks{X. Tu and H. S. Dhillon are with Wireless@VT, Bradley Department of Electrical and Computer Engineering, Virginia Tech, Blacksburg, VA, 24061, USA. Email: \{xiangliutu, hdhillon\}@vt.edu. C. Saha is with the
Qualcomm Standards and Industry Organization, Qualcomm Technologies
Inc., San Diego, CA 92121 USA. C. Saha contributed to this work during his affiliation with Virginia Tech. Email: csaha@vt.edu. The support of the US NSF (Grant CNS-1923807) is gratefully acknowledged.
}
}

\maketitle  
\begin{abstract}

Constructing a no-fly zone (NFZ) is a straightforward and effective way to facilitate the coexistence of unmanned aerial vehicles (drones) and existing systems (typically satellite systems). 
However, there has been little work on understanding the optimal design of such NFZs. 
In the absence of this design, one invariably ends up overestimating this region, hence significantly limiting the allowed airspace for the drones. 
To optimize the volume of the NFZ, we formulate this task as a variational problem and utilize the calculus of variations to rigorously obtain the NFZ as a function of the antenna pattern of victim receivers and the spatial distribution of drones. 
This approach parallels the matched filter design in the sense that the NFZ 
extends in directions where the antenna gain and/or the density of drones is high. 
Numerical simulations demonstrate the effectiveness of our optimal design compared to the known baselines in reducing the volume of the NFZ without compromising the protective performance.  
\end{abstract}
\section{Introduction}    

Enabling the coexistence of drones and existing communications systems is crucial for the efficient utilization of spectrum. 
Due to their large coverage footprints, drones can generate high aggregate interference to the ground-based infrastructure. 
Satellite systems with ground stations (GSs) pointing upwards are particularly susceptible to such interference even when the drones operate in the adjacent bands.
This scenario constitutes the main setting of interest in this paper.
A straightforward but largely effective way to safeguard such highly sensitive ground nodes is to create a {\em no-fly} zone (NFZ) around them. 
The NFZ is a designated geographic area where drone operations are either restricted or entirely prohibited. 
It essentially enforces a minimum required separation distance between drones and GSs. 
Regulatory bodies have provided initial guidance on this minimum separation distance for various systems. 
For instance, drones operating in the 1710-1785 MHz range can potentially interfere with Meteorological Satellite (MetSat) systems operating in the 1675-1710 MHz. 
Accordingly, the regulatory measures require a minimum separation distance of 30 km for drones operating on the first adjacent channel to MetSat systems~\cite{ECCREPORT309}.
While this idea has been discussed from the regulatory perspective, quite remarkably, there has been little work on understanding the {\em optimal design} of such NFZs. 
In the absence of this design, one invariably overestimates this zone, leading to unnecessary large NFZs that significantly limit the operational airspace for drones. 
For example, even with the $-40$ dBm/MHz emission limit for drones, the required minimum separation distance can extend to $1.4$ km. 
In practice, NFZs are typically implemented in the form of domes or cylinders, with the minimum separation distance determining their radius. 
With this separation distance extending over dozens of kilometers, the resulting NFZ imposes significant restrictions on the operational space for drones. 
The shapes of these NFZs, typically domes, are selected based on the assumption of a uniform interference risk around the victim receiver. 
However, this assumption is usually highly conservative given that satellite systems often use directional antennas. 
Consequently, the minimum separation distance can be reduced for drones positioned outside the main lobe of the GS antennas. 
This implies that the separation distance should be designed based on the spatial distribution of drones relative to the antenna pattern of the GSs. 
This approach is particularly appealing when the antenna pattern remains fixed for an extended time.
Generally, assuming that the separation distance varies continuously with the elevation and azimuth angle in 3D space, the NFZ is the region delineated by the minimum separation distance, as denoted by the red part in~\figref{networkillustrate}. 
In addition to spatial separation, spectral separation methods such as guard bands and frequency masks are employed to reduce interference between systems. 
By optimizing frequency and distance separation jointly, it is feasible to further reduce the volume of NFZs, which is also explored in this paper. 

The coexistence of drone and satellite systems has been extensively explored in the literature, e.g., see~\cite{kim2023feasibility,sharma2016line,cho2020coexistence,azari2017coexistence,li2020maritime}. 
The usual approach to facilitating this coexistence is through spatial and/or spectral separation, both of which are interesting to this paper. 
Spatial separation results in the formation of NFZ, also termed exclusion regions or guard zones. 
The improvement in performance through spatial separation has been studied recently in~\cite{choi2021analysis,vinogradov2019uav}, which investigates the minimum distance separation for satellite systems coexisting with drones, a problem of particular interest to us. 
Along the same lines, the authors of~\cite{kishk2017coexistence} have studied the use of circular secrecy guard zones to balance protecting the confidentiality of the primary users while ensuring that the secondary devices harvest sufficient energy from the radio frequency transmissions of the primary network.
Beyond communications-related interference, the idea of guard zones has also been applied to avoid physical collisions between drones and other aircraft or facilities, where it is termed geofencing~\cite{vagal2021new,cho2018assess}. 
On the spectral front, frequency masks and guard bands offer another layer of separation for achieving coexistence. 
For example, the analytical guard band is proposed in~\cite {demmer2018analytical} to limit the distortion caused by 5G multi-service multiplexing, while~\cite{lott1997calculation} studies the minimum frequency separation for the coexistence of mobile communication systems. 
In addition to these investigations, efforts are underway to address the challenges posed by the existence of NFZs, such as designing strategies for cooperative multi-drone area search~\cite{bo2015multi,xu2020multiuser} and optimizing routes for drone-based delivery and pickup~\cite{jeong2019truck} with NFZ constraints. 
Despite these studies, the optimal design of NFZs remains a notable open problem. 

{\em Contributions:}
In this paper, we investigate the optimal design of the NFZ to strike a careful trade-off between safeguarding satellite systems from drone interference while preserving adequate operational space for drones. 
The goal is to identify an optimal separation distance function that minimizes the volume of the NFZ while satisfying the regulatory measures for protecting the link between GS and satellites. 
Additionally, we investigate the use of guard bands and frequency masks for frequency separation in modeling drone interference. 
While expanding the guard band can reduce the separation distance, it concurrently decreases the bandwidth efficiency for drones which presents another trade-off explored in this paper. 
To address the optimal design problem, we approach it as a variational problem and use the calculus of variation to rigorously derive the optimal separation distance function. 
Furthermore, we establish interesting parallels between the optimal solution and matched filter design principle commonly used in communications systems. 
As expected, the numerical results show that the optimally designed NFZ is more effective than baseline shapes in reducing the volume of the NFZ while maintaining the same protective performance. 

\section{System Model}
\begin{figure}[hbpt]
    \centering
    \includegraphics[width=0.41\textwidth]{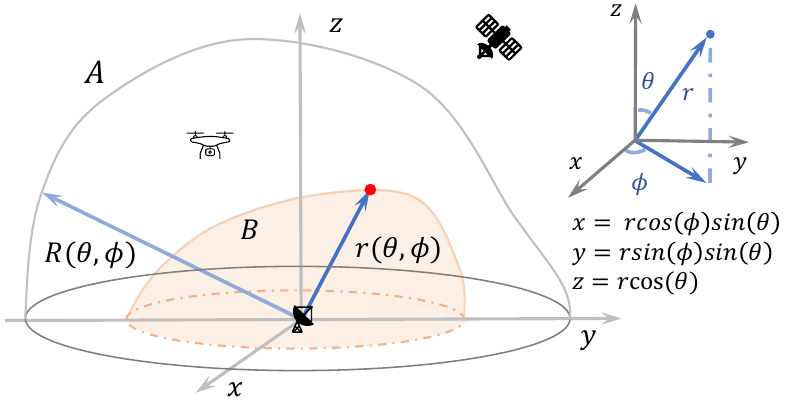} 
    \caption{ An illustration of the system model and NFZ. The red region denotes the NFZ $B$ and the red point indicates an arbitrary point on the NFZ boundary. }
    \label{networkillustrate}
\end{figure}

\begin{figure}[hbpt]
    \centering \includegraphics[width=0.4\textwidth]{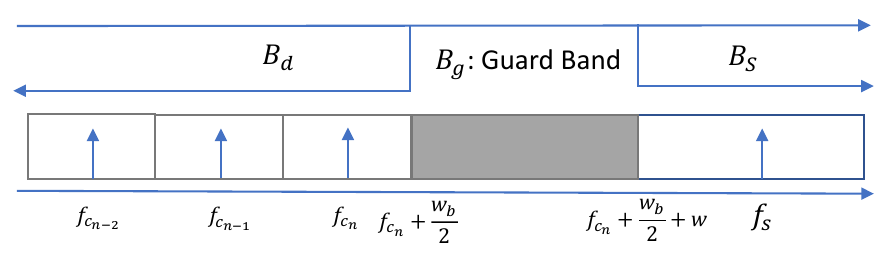} 
    \caption{ The illustration of frequency separation and the guard band.}
    \label{frequencyseperation}
\end{figure}

We consider the satellite network with a GS located at $\nbx_g=(x_g,y_g,z_g)$, coexisting with a 3D drone network. Drones situated within the region $A = \left\{ (x, y, z) \mid \sqrt{d_{x}^2 + d_{y}^2 + d_{z}^2} \leq R(\theta,\phi), z\geq 0 \right\}$ are considered to contribute to the interference experienced by the GS. 
Here, $d_x=x-x_g$, $d_y=y-y_g$ and $d_z=z-z_g$. $R(\theta,\phi)$ denotes the distance from the boundary of $A$ to the GS where $\theta\in [0, \frac{\pi}{2}]$ and $\phi\in [0, 2\pi]$ are the elevation and azimuth angle, respectively.
The NFZ $B$ is a subset of $A$ and defined as $B = \left\{(x,y,z) \mid \sqrt{d_x^2 + d_y^2 + d_z^2} \leq r(\theta, \phi), z\geq 0 \right\}$, where $r(\theta, \phi)$ represents the minimum separation distance between the edge of $B$ and GS. 
\figref{networkillustrate} illustrates the system model.
The drone locations follow a 3D inhomogeneous Poisson point process (PPP) $\Phi = \{\nbX_i,i\in \nbbN \}$ with intensity $\lambda(\nbx)$, where $\nbX_i = (x_i, y_i, z_i) $ denotes the location of $i$-th drone. 
Assume, quite realistically, that the GS has a directional antenna whose antenna gain is given by an arbitrary non-negative function $g(\theta, \phi)$. 
We assume the satellite link occupies a frequency band of bandwidth $B_{S}$ centered at $f_s$ MHz, whereas the drones operate in its adjacent band with bandwidth $B_{D}$ centered at $f_{d}$ MHz. The band $B_{D}$ is divided into two parts: the communication band $B_{d}$, which is exclusively used for drone communications, and the guard band $B_g$ with bandwidth $w$ MHz as illustrated in~\figref{frequencyseperation}. 
The band $B_d$ is further divided into $K\in \nbbN^+$ resource blocks, each with a bandwidth $w_b = \frac {B_{d}}{K}$.  
The center of the $k$-th block is denoted by $f_{c_k}$.
Each drone selects one of the $K$ resource blocks uniformly at random.  
Therefore, the drones using the $k$-th resource blocks form an inhomogeneous PPP $\Phi_k$ with intensity $\lambda_k(\nbx) = \lambda(\nbx)/K$. 
The guard band $B_g$ is located between $B_{d}$ and $B_{s}$. 
Expanding the guard band effectively mitigates interference to GS, consequently releasing operational space for drones. However, this comes at the expense of reduced spectral efficiency. Therefore, balancing the operational space and frequency availability is critical when implementing the guard band. 
To analyze this trade-off, we first define the unwanted emission generated by the set of drones using $k$-th resource block 
  \begin{align}
      P_k = \int_{f_{c_k}+\frac{w_b}{2}+(N-k)w_b+w}^{f_{c_k}+\frac{w_b}{2}+(N-k)w_b+w+B_s} P_t H(f) \, \nrmd f, 
      \label{p_k}
  \end{align}
where $H(f)$ is the emission mask of the drone normalized by the transmit power $P_t$ of the drone. 
We assume each drone is equipped with an omnidirectional antenna. 
Therefore, the interference power received at the GS caused by drones operating in the $k$-th resource block is $I_k = \sum_{\nbx_i \in \Phi_k \cap A \backslash B } P_k G_i L_i,$
and the total interference is $I  = \sum_{k=1}^{K} I_k$, where $G_i = g(\theta_i, \phi_i)$ is the GS antenna gain, and $\theta_i $ and $\phi_i$ are the elevation and azimuth
the angle of $i$-th drone, respectively. 
Here, $L_i = l(\|\nbx_i-\nbx_g\|)$ is the general path loss dependent only on the distance between drones and the GS.

\subsection{Problem Formulation}
Our goal is to determine the optimal separation distance function $r(\theta,\phi)$ which allows us to construct a NFZ with minimum volume while effectively protecting GS to reduce the interference to comply with the regulatory requirement.
For this, we characterize the expectation of total interference $\nbbE_{\Phi}[I]$ which can be mathematically expressed as
\begin{align}
    \nbbE_{\Phi}\!\left[I\right] =& \, \nbbE_{\Phi}\!\bigg[ \sum_{k=1}^{K} I_k\bigg]
    =\sum_{k=1}^{K} \nbbE_{\Phi_k}\!\left[ \sum_{\nbx_i \in \Phi_k \cap A \backslash B } P_k G_i L_i\right]  \\
    \overset{(a)}{=}&   \bigg[ \int_{A \backslash B}   \lambda_k\!\left(\nbx\right)  \sum_{k=1}^{K}   P_k \, l\!\left(\|\nbx \|\right) g\!\left(\nbx\right)  \nrmd \nbx \bigg] \\ 
    =&   \int_{0}^{2 \pi} \int_{0}^{\frac{\pi}{2}} \int_{0}^{R(\theta, \phi)}    P \lambda   l\!\left(\rho\right) g  \rho^2 \sin\!\theta\, \nrmd \rho \nrmd \theta \nrmd \phi \notag  \\ 
    & - \int_{0}^{2 \pi} \int_{0}^{\frac{\pi}{2}} \int_{0}^{r} P \lambda  l\!\left(\rho\right) g  \rho^2\sin\!\theta \, \nrmd \rho \nrmd \theta \nrmd \phi  \label{eq:E[I]} \\
    = &\nbbE_{\Phi}\!\left[I_{A}\right] -\nbbE_{\Phi} \!\left[I_{B}\right],
\end{align}
where we denote $\lambda(\theta,\phi,\rho)$ by $\lambda$, $g\!\left(\theta,\phi\right)$ by $g$ and $r(\theta,\phi)$ by $r$  for notational simplicity. 
Step $(a)$ follows by applying Campbell's theorem~\cite{haenggi2012stochastic}.
$P = \frac{1}{K}\sum_{k=1}^{K} P_k$ is the average power over $K$ resource blocks. 
The equation~\eqref{eq:E[I]} can be interpreted as follows: the first integral represents the expected total interference received at GS without implementing a NFZ, denoted as $\nbbE_{\Phi}\!\left[I_{A}\right]$, and the second integral denotes the expected interference reduction by removing the drones from the NFZ $B$, denoted as $\nbbE_{\Phi} \!\left[I_{B}\right]$. 

To protect the GS from drone interference, we use the following regulatory constraint: $\nbbE_{\Phi}\!\left[I\right] < a'$. 
Note that even though we are placing a constraint on the mean interference here, it implicitly bounds the tail of the interference distribution through Markov's inequality: $P(I \geq \beta) \leq E_\Phi[I] / \beta$. 
Given a fixed region $A$, $\nbbE_{\Phi}\!\left[I_{A}\right]$ is constant for specific path loss, g, and $\lambda$. Consequently, the interference mitigated by constructing the NFZ $B$ needs to satisfy: 
\begin{align}
 \nbbE_{\Phi} \!\left[I_B\right] =\int_{0}^{2 \pi} \int_{0}^{\frac{\pi}{2}} \int_{0}^{r} P \lambda  l\!\left(\rho\right) g  \rho^2\sin\!\theta \, \nrmd \rho \nrmd \theta \nrmd \phi \geq a , 
\label{eq:E(I_e)}
\end{align}
where $a = \nbbE_{\Phi}\!\left[I_{A}\right]-a'$. 
As the integrand is non-negative, achieving a higher interference reduction $\nbbE_{\Phi} \!\left[I_{B}\right]$ requires expanding the NFZ. 
Therefore, implementing the NFZ poses a challenge rooted in the trade-off between two considerations: expanding the separation distance $r$ for effective interference reduction (interference mitigation) while ensuring sufficient operational space for drones (operational flexibility). 
Building on this, the problem is formally characterized as follows:
\begin{subequations}
\begin{align}
\label{eq:opt:objective}
\min_{r(\theta, \phi)} & \, \int_{0}^{2 \pi} \!\int_{0}^{\frac{\pi}{2}}\! \int_{0}^{r(\theta, \phi)} \!\rho^2 \sin\!\theta \, \nrmd \rho \nrmd \theta \nrmd \phi \\
\label{eq:opt:constriant} 
{\rm s.t.}& \, \, \nbbE_{\Phi} \!\left[I_B\right]  \geq a.
\end{align}
\label{eq:opt:optimization}
\end{subequations}
It is easy to recognize that the problem~\eqref{eq:opt:optimization} falls into the calculus of variations domain, as both the objective and constraint are functionals dependents on the function $r(\theta,\phi)$.

\section{Optimal Design of No-fly Zone} 
In this section, we will analyze the solution to problem~\eqref{eq:opt:optimization} while introducing fundamental concepts of the calculus of variations. 
Interested readers can refer to~\cite{kot2014first} for a more systematic treatment of this topic. 
The roots of the calculus of variations trace back to the well-known brachistochrone problem posed by Johann Bernoulli in 1696~\cite{Johann1696}. It finds the path of fastest descent between two points under gravity, which is a cycloid. 
In general, the calculus of variations deals with optimizing the functional, which is a mapping from a set of functions to real numbers. More formally, function~\eqref{eq:opt:objective} and~\eqref{eq:opt:constriant} can be expressed as a functional of $r(\theta, \phi)$:
\begin{align}
     J[r] &= \int_{0}^{2\pi}\int_{0}^{\frac{\pi}{2}}\int_{0}^{r} \rho^2 \sin\!\theta \, \nrmd \rho \nrmd \theta \nrmd \phi, \label{eq:functional:objective}\\
     K[r] &= \!\int_{0}^{2 \pi} \! \int_{0}^{\frac{\pi}{2}} \! \int_{0}^{r}\! P \lambda l\!\left(\rho\right)\! g  \rho^2 \sin\!\theta\, \nrmd \rho \nrmd \theta \nrmd \phi -  a=0,  \label{eq:functinoal_constraint}
\end{align}
where we substitute the inequality~\eqref{eq:opt:constriant} with equality. As noted above, eliminating larger interference necessitates expanding NFZ; therefore, the NFZ with the smallest volume must satisfy $\nbbE_{\Phi} \!\left[I_{B}\right]  = a$.
Now, the problem of interest is to find the optimal $r^{*}$ that minimizes $J[r]$ while $K[r^*]=a$.
To obtain the solution, we first modify the $r^{*}$ by introducing a variation $\delta r$, defined as
\begin{align}
    \delta r(\theta,\phi) = \epsilon \eta(\theta,\phi) =  r(\theta,\phi) -r^*(\theta,\phi), 
\end{align}
where $\epsilon$ is a scalar and $\eta(\theta,\phi)$ is an arbitrary small weak variation.
Therefore, the functional~\eqref{eq:functional:objective} changes to 
\begin{equation}
    \label{eq:changed-functional}
    J[r] = J[r^* + \delta r] = \int_{0}^{2\pi}\int_{0}^{\frac{\pi}{2}}\int_{0}^{r^* + \delta r} \rho^2 \sin\!\theta \, \nrmd \rho \nrmd \theta \nrmd \phi.
\end{equation}
Subtracting~\eqref{eq:functional:objective} from~\eqref{eq:changed-functional}, the increment of functional $J$ is
\begin{align}
    \Delta J &= J[r] - J[r^{*}] = J[r^* + \delta r] - J[r^*] .
\end{align}
Following this, we define the auxiliary functional of $J[r]$ and $K[r]$ as 
$G[r] = J[r] + \mu K[r] $, where $\mu \in \nbbR$ is the Lagrange multiplier.
The first variation of the $G[r]$ is defined as
\begin{align}
    \delta G &= \lim_{\epsilon \rightarrow 0} \frac{\nrmd G}{\nrmd\epsilon} = \lim_{\epsilon \rightarrow 0} \frac{G[r^* + \delta r] - G[r^*]}{\epsilon}  \\
    &= \frac{\nrmd G[r^* + \delta r]}{\nrmd \epsilon} \bigg|_{\epsilon=0} \overset{(a)}{=}  \frac{\partial J}{\partial r} \frac{\partial r}{\partial \epsilon} + \mu \frac{\partial K}{\partial r} \frac{\partial r}{\partial \epsilon} \\
    &\overset{(b)}{=} \int_{0}^{2\pi}\int_{0}^{\frac{\pi}{2}}  \left[F_r +\mu W_r \right] \eta(\theta,\phi) \,\nrmd \theta \nrmd \phi,
\end{align}
where $F_r = \frac{\partial F(r(\theta,\phi))}{\partial r(\theta, \phi)}  $ and 
 $W_r = \frac{\partial W(r(\theta,\phi))}{\partial r(\theta, \phi)} $. 
The function
$F = \frac{1}{3}r(\theta,\phi)^3\sin\!\theta$ and $W = P U(\theta,\phi,r)g(\theta,\phi) \sin\!\theta-\frac{a}{\pi^2}$
are the integrands obtained by taking the integral of equation~\eqref{eq:functional:objective} and~\eqref{eq:functinoal_constraint} with respect to $\rho$. 
Step $(b)$ is followed by the Leibniz integral rule, where the order of partial derivation and integral are exchangeable.
\begin{remark}
With the linearization of the functional, the first variation of $G$ can be used as an approximation of the increment $\Delta G = \delta G + o\!\left(d_1\!\left(r, r + \epsilon \eta\right)\right)$, where $d_1(r_1, r_2)$ is the first order distance between two functions. 
\end{remark}
Now, we present the necessary conditions for $G[r]$ to take an extremum in Lemma~\ref{lemma1}. Please refer to~\cite{kot2014first} for its proof. 
\begin{lemma}
A necessary condition for the functional $G[r]$ to have 
local extremum at $ r(\theta,\phi)= r^{*}(\theta,\phi) \in C^{\infty}[0,2\pi]\times[0,\frac{\pi}{2}]$ is that the first variation of the functional $\delta G[r^*]= 0 $, for all admissible variation $\eta(\theta,\phi)$. 
\label{lemma1}
\end{lemma}
The fundamental lemma of the calculus of variations is given below. Please refer to~\cite{kot2014first} for its proof.
\begin{lemma}
 If $S(x)$ is continuous on $[a,b]$ and $\int_a^b S(x)\eta(x)\, \nrmd x = 0$, for all compactly supported smooth variation $\eta(x)$, then $S(x) = 0, \forall \, x\in [a,b]$.
\label{lemma2}
\end{lemma}
Drawing on Lemmas~\ref{lemma1} and~\ref{lemma2}, we present the Euler-Lagrange condition for the optimal distance $r^*(\theta,\phi)$ that enables $J[r]$ to achieve its extremum in the following theorem.
\begin{thm}
    Given an arbitrary antenna pattern $g(\theta,\phi)$, a path loss model $l(\cdot)$, and the spatial distribution of drone $\lambda(\theta,\phi,\rho)$, the NFZ with minimum volume is constructed by the separation distance $r^*$ that satisfies the following
    \begin{align}
        P \lambda(\theta,\phi,r^*(\theta,\phi))l(r^*(\theta,\phi)) g(\theta,\phi) = \mu,
        \label{eq:them:distance}
    \end{align} 
    \label{eq:them}
    where $\mu \in \nbbR$ is the Lagrange multiplier.
\end{thm}

\begin{IEEEproof}
From Lemmas~\ref{lemma1} and~\ref{lemma2}, the necessary condition for the objective functional $J[r]$ to achieve the extremum  under the constraints $K[r]$ is that there exists $ \mu $ such that
\begin{align}
\frac{\partial F(r(\theta,\phi))}{\partial r(\theta, \phi)}   + \mu \frac{\partial W(\theta,\phi)}{\partial r(\theta, \phi)} = 0.
\end{align}
By taking the partial derivative of function $F$ and $W$,
we have
\begin{align}
      1+\mu P \lambda(\theta,\phi,r(\theta,\phi)) l(r(\theta,\phi)) g(\theta,\phi) = 0.
      \label{eq:them:proof_fg}
\end{align}
Since $\mu \in \nbbR$ is a constant, we can get the result in~\eqref{eq:them:distance}. 
\end{IEEEproof}

\begin{cor}
    If the drones form a homogeneous PPP with constant density $\lambda$,  the NFZ with minimum volume is constructed by the separation distance given by
    \begin{align}
        r^*(\theta,\phi) = l^{-1}\!\left(\frac{\mu}{\lambda P g(\theta,\phi)} \right)  ,
        \label{eq:them:distance_homo}
    \end{align} 
    where $l^{-1}$ is the inverse of the path loss function.
\end{cor}
\begin{IEEEproof}
From~\eqref{eq:them:proof_fg}, we have $\lambda P l(r(\theta,\phi)) g(\theta,\phi) = \mu. $
Solving for  $r(\theta,\phi)$, we get~\eqref{eq:them:distance_homo}. 
\end{IEEEproof}
To demonstrate the performance of the optimal separation distance $r^*(\theta,\phi)$, we will incorporate specific pathloss functions and antenna patterns into our framework.  
\vspace{-4mm}
\subsection{Path loss model and antenna pattern}
\begin{figure*}[hpt]
    \centering
    \subfigure[Interference elimination of three NFZ shapes.]{ 
        \begin{minipage}[t]{0.29\textwidth}
        \centering
        \includegraphics[width=\textwidth]{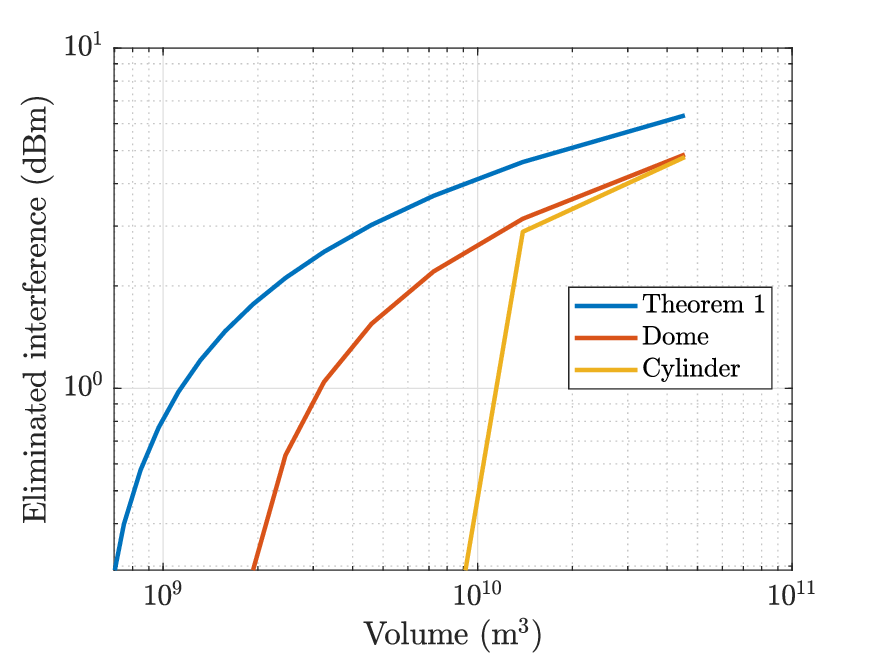}
        \label{interference}  
        \end{minipage}}
        \subfigure[Sliced view of NFZ with $\phi=0$.]{
        \begin{minipage}[t]{0.29\textwidth}
        \centering
        \includegraphics[width=\textwidth]{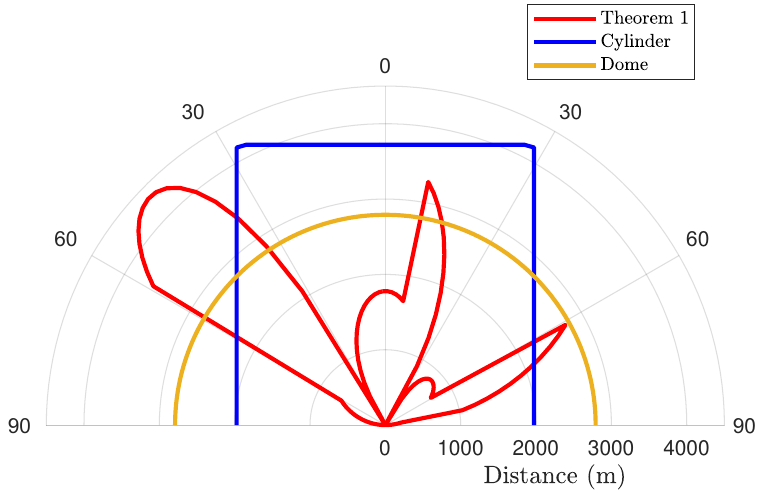}
        \label{cutview}  
        \end{minipage}}
        \subfigure[Sliced view of the drone distribution density.]{
        \begin{minipage}[t]{0.29\textwidth}
        \centering
        \includegraphics[width=\textwidth]{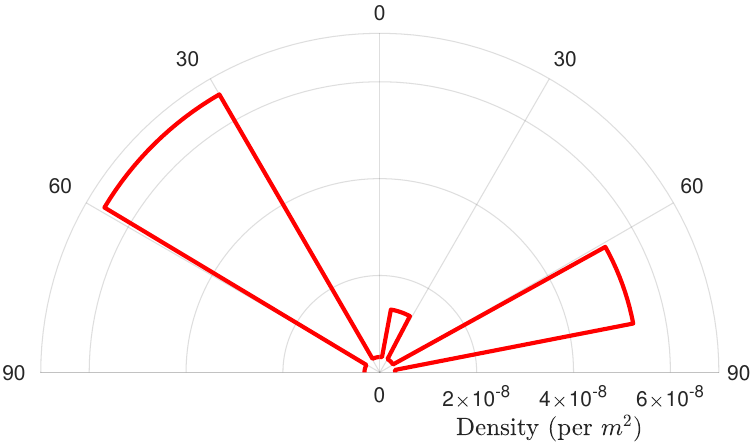}
        \label{density}  
        \end{minipage}}
    \subfigure[Antenna pattern $g(\theta,\phi)$ of BS.]{ 
        \begin{minipage}[t]{0.29\textwidth}
        \centering
        \includegraphics[width=\textwidth]{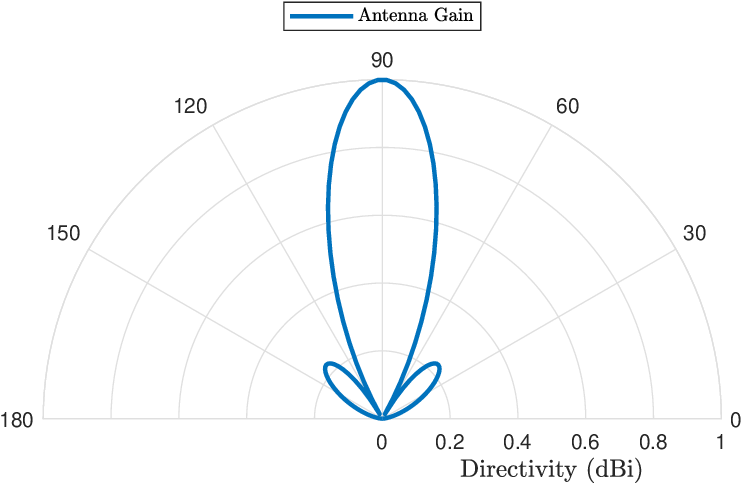}
        \label{antenna_pattern}
        \end{minipage}}
        \subfigure[Emission mask $H(f)$ of drone.]{
        \begin{minipage}[t]{0.29\textwidth}
        \centering
        \includegraphics[width=\textwidth]{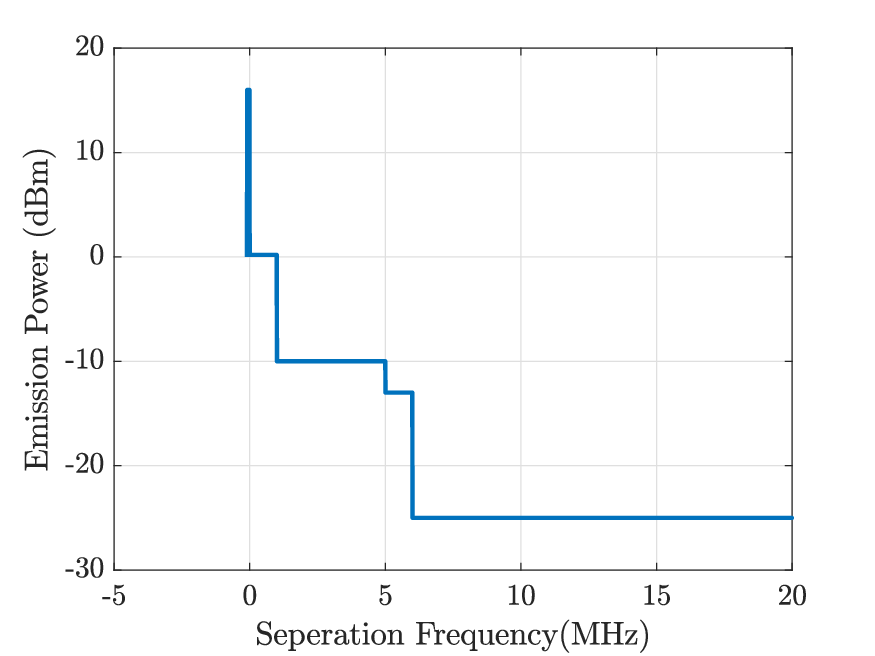}
        \label{emission_mask}
        \end{minipage}}
        \subfigure[Eliminated interference for different guard band.]{
        \begin{minipage}[t]{0.29\textwidth}
        \centering
        \includegraphics[width=\textwidth]{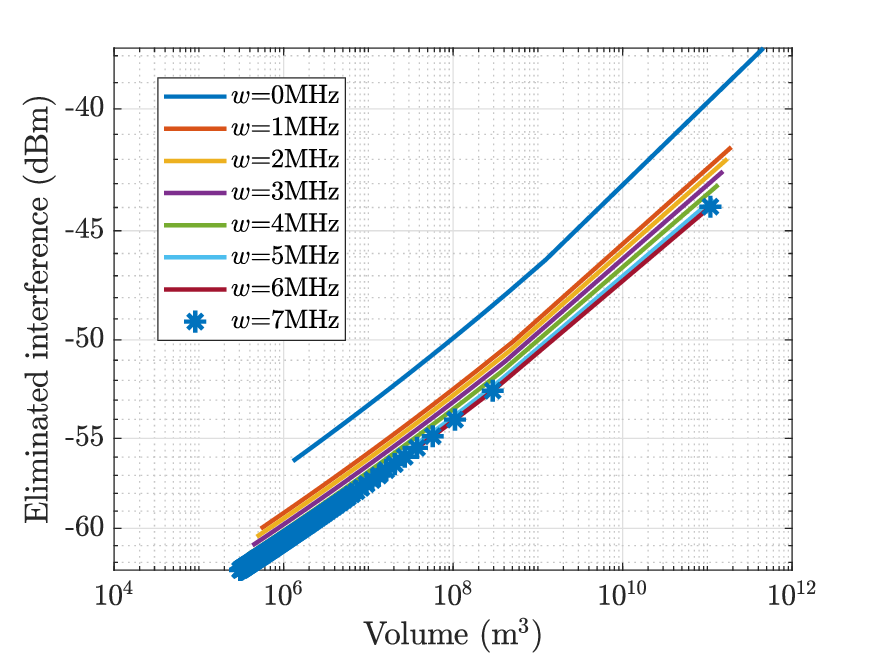}
        \label{seperation_frequency}
        \end{minipage}}
        \caption{ (a) shows the eliminated interference for three shapes of NFZ. (b) (c) (d) are the sliced views of the NFZs, the distribution density of drones and the antenna pattern of GS. (e) is the emission mask of drones. (f) shows the reduction of NFZ volume as the guard band expands. } 
 \end{figure*}
To maintain realism, we use a bounded power-law path loss: $l(r)=\min \left\{1,r^{-\alpha}\right\}$, where $r$ is the distance between GS and drones and $\alpha > 0$ is the path loss exponent. 
For the antenna pattern, consider a uniformly linear array with $M$ elements
\begin{align}
    \label{eq:ULA-pattern}
    g(\theta,\phi) =\left| \frac{\sin\!\left( M\frac{\pi d }{\lambda_f} \sin\!{\theta} \right)}{\sin\!\left( \frac{\pi d }{\lambda_f} \sin\!{\theta} \right)}\right|,
\end{align}
where $d$ is the spacing between antenna elements, and $\lambda_f$ is the wavelength.
Using the conclusion in Theorem~\ref{eq:them}, the optimal separation distance is given as
\begin{align}
    r^*(\theta,\phi) = \max\!\left\{1, \left(\frac{P\lambda(\theta,\phi)g(\theta,\phi) }{\mu}  \right)^{\frac{1}{\alpha}}\right\}.
    \label{eq:opt:distance}
\end{align}
The equation~\eqref{eq:opt:distance} reveals that the optimal separation distance $r^*(\theta,\phi)$ is proportional to $\left(\lambda(\theta,\phi)g(\theta,\phi)\right)^{1/\alpha}$. 
{\it This can be interpreted as the spatial matched filter, which expands the NFZ in the directions where the antenna gain and/or the spatial density of drones is high.}
\subsection{Results and Discussion }
The simulation parameters are set as follows: $M=8$, $d / \lambda_f = 0.25$ and $\alpha = 2.5$. 
Without loss of generality, we assume that the GS is located at $(0, 0, 0)$ and the main lobe of the GS antenna is pointing straight up.
This setup can easily be generalized to other settings since there are no such restrictions in our analytical results. 
To evaluate the performance of the optimal NFZ, we compare the ability to eliminate interference against two baseline shapes: the dome and the cylinder, with the same volume.
The dome-shaped NFZ is a reasonable canonical model and is especially meaningful when no information is available about the antenna patterns and/or the spatial distribution of the drones. 
The cylinder is another meaningful shape since the main lobe of the GS always points in a specific direction. 

{\em Non-homogeneous PPP:}
In real networks, the distribution of drones is not necessarily homogeneous across the entire space since they always fly within a specific altitude range. For our representative results, we set 
\begin{equation}
     \lambda = \left \{
     \begin{aligned}
        & 10^{-8}   , &  10^o\leq \theta \leq 30^o,\, 0^o\leq \phi \leq 360^o ,\\
        & 10^{-7.2} , & 30^o  < \theta \leq 60^o,\, 0^o\leq \phi \leq 180^o , \\
        & 10^{-7.3} , & 60^o  < \theta \leq 80^o,\, 180^o\leq \phi \leq 360^o ,\\
        & 10^{-8.5} , & \text{otherwise}. 
     \end{aligned} 
     \right. 
\end{equation} 
The ability to eliminate interference for different shapes is depicted in~\figref{interference}. 
The simulation results show that the optimal NFZ requires the smallest volume to eliminate the same interference. 
For cylindrical NFZ, we simulate to find the radius-height ratio that eliminates most interference. 
To illustrate the matched filter design aspect of our method, we provide slices of these 3D NFZs for $\phi=0, \pi$ in~\figref{cutview}.  
When comparing these slices with the antenna pattern $g(\theta,\phi)$ and the density $\lambda(\theta,\phi)$ of drones depicted in~\figref{antenna_pattern} and~\figref{density}, it is evident that the optimal NFZ matches closely to the shape of $g(\theta,\phi)$ and $\lambda(\theta,\phi)$. The optimal NFZ effectively prevents interference from drones by expanding the separation distance in areas with higher antenna gain and drone density and reducing the separation distance in the other directions. 

{\em Guard band:}
In modeling the unwanted emission power of drones, we incorporate the frequency mask shown in~\figref{emission_mask}.
Each resource block possesses $w_b = 5$ MHz bandwidth. 
We plot the eliminated interference of the optimal NFZ with guard bands $w = \{0,\cdots,7\} $ MHz in~\figref{seperation_frequency} which show a reduction in the volume of the NFZ with the expansion of guard band $w$. 
The most substantial decrease in the volume occurs when the guard band is $1$ MHz which corresponds with the emission mask characteristics where the emission power experiences a notable decrease at a frequency separation of $1$ MHz. 
According to equation~\eqref{p_k}, implementing guard band changes the unwanted emission $P_k$ and hence $\nbbE_{\Phi}\!\left[I_B\right]$. 
Therefore, joint tuning of the guard band and NFZ is necessary to obtain the desired outcome for a given application scenario.

\section{Conlusion}
The NFZ is a straightforward and effective way to protect sensitive nodes from interference. To strike a careful balance between having a large enough NFZ to offer protection to the victim receivers and providing enough operational space to the coexisting drones, we formulated and solved the problem of optimal NFZ construction using the calculus of variation.  
The optimal separation distance function parallels the matched filter design where the NFZ is larger in the direction of high antenna pattern and/or drone density. 
Our analysis also provided crucial insights into the tradeoff between spatial and spectral separation in this coexistence setting. 

\bibliographystyle{IEEEtran}
\bibliography{hokie}
\end{document}

%% file: notation.tex
\def\nbx{{\mathbf{x}}}

\def\nb0{{\mathbf{0}}}
\def\nb1{{\mathbf{1}}}

\def\nbX{{\mathbf{X}}}

\def\nbbE{{\mathbb{E}}}

\def\nbbN{{\mathbb{N}}}

\def\nbbR{{\mathbb{R}}}

\def\nrmd{{\rm d}}

\newtheorem{lemma}{Lemma}
\newtheorem{thm}{Theorem}

\newtheorem{cor}{Corollary}

\newtheorem{remark}{Remark}

\def\figref#1{Fig.\,\ref{#1}}%

